\documentclass[runningheads]{comsis2}
\pdfoutput=1

\usepackage{mathptmx}
\usepackage{graphicx}
\usepackage{times}
\usepackage{url}
\usepackage{subfigure}
\usepackage{fancyvrb}
\usepackage{multirow}
\usepackage{hhline}
\usepackage[labelsep=space]{caption}
\usepackage[usenames]{color}
\usepackage{graphicx}
\usepackage{times}
\usepackage{tipa}
\usepackage{url}
\usepackage[hidelinks]{hyperref} % latest version of MiKTex does not support the "hidelinks" option
\usepackage{cite}
\usepackage[parfill]{parskip}

\emergencystretch=\maxdimen
\hyphenchar\font=-1

\def\journalissue{}
\def\paperidnum{}
\begin{document}
\pagenumbering{arabic}
  \pagestyle{plain}

\title{Enhancing Programming Interface to Effectively Meet Multiple Information Needs of Developers}

% author names and affiliations use a multiple column layout for up to three different affiliations
\author{Haipeng Cai\inst{1}}
\institute{Department of Computer Science and Engineering, University of Notre Dame\\
  Notre Dame, IN, 46556, USA\\
  \email{hcai@nd.edu}
}

\maketitle
\thispagestyle{plain}

\begin{abstract}
%\vspace{-3pt}
In the past decades, integrated development environments (IDEs) have been largely advanced to facilitate common software engineering tasks. Yet, with growing information needs driven by increasing complexity in developing modern high-quality software, developers often need to switch among multiple user interfaces, even across different applications, in their
development process, which breaks their mental workflow thus tends to adversely affect their working efficiency
and productivity.

This position paper discusses challenges faced by current IDE designs mainly from working context transitions of developers
during the process of seeking multiple information needs for their development tasks. It remarks the primary
blockades behind and initially explores some high-level design considerations for overcoming such challenges in the next-generation IDEs.
%
%Inspired by useful interface and visualization design features in visual programming environments,
Specifically, a few design enhancements on top of modern IDEs are envisioned, attempting to
reduce the overheads of frequent context switching commonly seen in the multitasking of developers.
%seen by developers when seeking various information they need.
%In this paper, we highlight the problem with working context transitions in existing IDEs, and remark the primary blockades behind; we then briefly describe the high-level design considerations for overcoming those blockades in the next-generation IDEs.%; finally, %we rough out the preliminary agenda regarding time schedule and management approaches concerning research facilities.
%we line out the time schedule for the project and summarize the our solution to the present problem.

\vspace{6pt}\textbf{Keywords:} Information need, integrated development environment, context switching, automatic recommendation, programming interface, software visualization
\end{abstract}

\section{Introduction}
%\vspace{-2pt}
%Information can be more effectively grasped from data that are graphically represented than from those that
%are presented as they are originally produced. As information visualization makes data, especially those of large scale and/or complexity, more informative~\cite{mackinlay1986automating}, it has
%been introduced to computer software development to enrich the expressiveness of programming languages and development environments~\cite{mackinlay2007show,bostock2009protovis}. One instance of this particular application of graphical representations is visual programming~\cite{Metoyer2012UVL,CaiCAL13}, which has been shown to be an efficacious means for improving software development productivity, for non-professional programmers in particular.

One merit with visual programming~\cite{green1996usability,Metoyer2012UVL,CaiCAL13} is that its integrated interface empowers smooth transitions among the workflow steps of developers---the interface provides all programming elements (of visual forms) so that the developers involved in the interface can easily maintain their mental workflow models by focusing on mostly just one type of interface (i.e., visual).

With most existing IDEs (e.g., ECLIPSE~\cite{eclipse}), however, developers often face challenges from frequent
transitions between coding (text interface) and visual aids (graphical interface), or even between disparate applications (and their different interfaces)~\cite{lawrance2013programmers}. Since traditional (textual) programming involves typically a demanding logic reasoning process,
such transitions and context switches can cause great inefficiency~\cite{czerwinski2004diary} in the development activity, even greater risks to the quality of resulting software.

The reason underneath is that context switching tends to interrupt the workflow~\cite{czerwinski2004diary} of developers.
More important, this problem can be even exacerbated by the growing information needs for developing modern software of increasing scale and complexity. Unfortunately, on the other hand, modern IDEs tend to grow in the complexity of their interface in a way that, seemingly facilitating developers to meet their needs for multiple sources of information, actually compounds the
problems with switching among increasingly more contexts.

%sequence of information-accessing activities,
%as against the software development process, the coding phase in particular, as a typical work flow, program analysis can be readily benefited from
%the power of information visualization as well. In actuality, program analysis often involves demanding logic reasoning, thus
%the graphical representations of abstract information are conducive to navigating the developers to get across complex relations
%among the large scope of elements in the working environment.

As it stands, research on easing programming tasks through interactive graphical environments
exists~\cite{Benton2007ISD} with most
focusing on providing visual aids within IDEs.
For instance, Dragon~\cite{chapman2003dragon} shows visual windows for program dependence, debugging information, data structure state, memory layout and similar other visual gadgets during program analysis tasks of developments, incorporated into the entire
software development work flow. However, this framework is limited to passively responding to user requests---it fails to automatically \emph{push} information to assist program-analysis tasks as if it were an integral part of the full task pipeline.
Thus, a more useful interactive programming environment needs to deliver informative visual aids not only
on demand but in a proactive manner, so as to minimize context switches during the entire development workflow.

Another challenge to today's IDE interfaces lies at their falling short of meeting the growing amount and \emph{variety} of information needs by developers. To finish a coding task, for example, developers usually have to consult many information sources that are diverse and distributed across disparate interfaces even applications. Although modern IDEs mostly have strong supports
to integrate diverse functionalities by means of plug-ins or extensions, information from those extraneous modules often has to
be passively retrieved under the requests of developers---the multiple sources of information are available, yet not well integrated \emph{in synergy}~\cite{zeller2007future} with other elements of the IDEs such that developers can
concentrate on their holistic workflow.

To address these specific issues,
this paper preliminarily explores several novel IDE features that could effectively assist
developers with handling multiple tasks while minimizing the costs of context switching during their development workflow. %smoothly transit between
%the visual interface, where visualizations of various types of assistant information are hosted, and the development arena, which is mainly the programming windows where developers design algorithmic procedures.
%To offer such instrumental features to the holistic IDE frameworks today, we borrow the general guidelines from visual programming environment design and the best practices of common graphical representation and human-computer interface design.
Specifically, three features are proposed focusing on interface design: extending traditional coding view to include 
co-worker views, offering automatic recommendation based information for API usage and code examples, and providing 
in-situ mechanism for mostly common used code-editing related operations driven by current the task context. And two major 
visualization features are discussed, including multiple code visualization views and interactive linked visualizations.
By illustrating the needs and benefits of these instrumental features, the paper demonstrates how
the next-generation IDEs could be
designed to offer better aids to developers in ways that improve development efficiency and productivity.

%This paper makes the following major contributions:
%\begin{itemize}
%\item We identify the context-switching issue in the design of today's IDEs that hinders the effectiveness of using them, and illustrate such issue using example usage scenarios.% (Section~\ref{sec:motivation}).
%%\item We propose overcoming such issue with an enhanced interactive programming framework that better meets various
%%    information needs of developers to performance multiples tasks at the same time in their development workflow (Section~\ref{sec:approach}).
%%\item We discuss several design improvements from interaction and visualization perspectives for the proposed framework, and illustrate how the new design features would work within existing IDEs to achieve the goals of the framework (Section~\ref{sec:interface} and~\ref{sec:vis}).
%\item We propose three interface design features that help reduce developers' overall cost of switching among multiple contexts in search of various sources of information.
%\item We propose two interactive visualization features that enable holistic integration of multiple
%    information in synergy so as to reduce developers' need of switching contexts when searching for various information.
%\end{itemize}
In summary, this \emph{position} paper highlights the context-switching issue in the design of today's IDEs that hinders the effectiveness of using them, and illustrate such issue using example usage scenarios; it discusses three interface design features that potentially reduce developers' overall cost of switching among multiple contexts in search of various sources of information; it also envisions two interactive visualization features that enable holistic integration of multiple
information in synergy so as to reduce developers' need of switching contexts when searching for various information.

The rest of this paper is organized as follows. First, Section~\ref{sec:motivation} gives a development scenario 
regarding information foraging that motivates our programming interface design. Then, Section~\ref{sec:interface} and 
Section~\ref{sec:vis} summarize the concrete features in the new programming environments, on interface and visualization design, 
respectively. Finally, Section~\ref{sec:concl} outlines the next step, planning on the implementation and evaluation of the 
proposed design.

\section{Motivating Example}\label{sec:motivation}
%\vspace{-2pt}
During software development, programmers gain most of the information they need from the source code they are working on. Yet,
they also need information beyond that, such as those produced by program analysis tools, to obtain better understanding of the software. Examples of such additional information include call graphs, program dependencies, and type hierarchies.
%dead code, performance optimization suggestions, and so on.
While most present IDEs do provide functionalities to help developers obtain these information, they force developers to
actively make requests for them. However, responding to user requests may not be sufficient in many situations. Rather,
a more effective IDE should provide developers with a \emph{guiding} interface instead of question responder, as developers may
not have \emph{prerequisite} information for them to initiate those requests or to do so in the most efficient way overall.
In consequence, excessive context switches ensue when developers have to resort to other contexts or even applications for
obtaining those missing information.
%
%What this implies is that just providing postmortem information is not an effective way of aiding the process of program analysis---users need immediate information to
%help them make decisions, which they cannot do before the present step is finished. That is, what the users really ask for is
%a guiding interface rather than a question responder.

In a typical usage scenario, a developer wants to know the overall design of the component-level architecture of a software for which he just finished the coding for one of its many packages. With a program analysis tool integrated in the IDE he is using, the developer \emph{has to} choose a button or menu item relevant to the functionality on the call graph of the entire program. Further, the developer proceeds by looking for all possible interfaces compatible for a function call of interest. Thus the developer has to traverse the call graph and hover mouse cursor over all relevant modules one by one.

However, without rich experiences with the very details of this software, it is infeasible for the developer to know how to make the preceding requests. The key issue, which is really the main obstacle here, is the requirement for the user to recognize which requests to make without auxiliary information provided by the program analysis tool. As such, the value of this visual-aid tool itself apparently diminishes. Arguably there exists a crucial need of developers with an IDE that incorporates interactive program analysis tools is a workflow-driven pipeline where the transitions from graphical to textual settings, and of course the other way around, are as seamless as possible. We are motivated by such an observation and the consequent requirement in the design of interactive programming interfaces.

\section{Interface Design}\label{sec:interface}
\begin{figure*}[htbp]
  \vspace{0pt}
  \centering
  \includegraphics[scale=0.85]{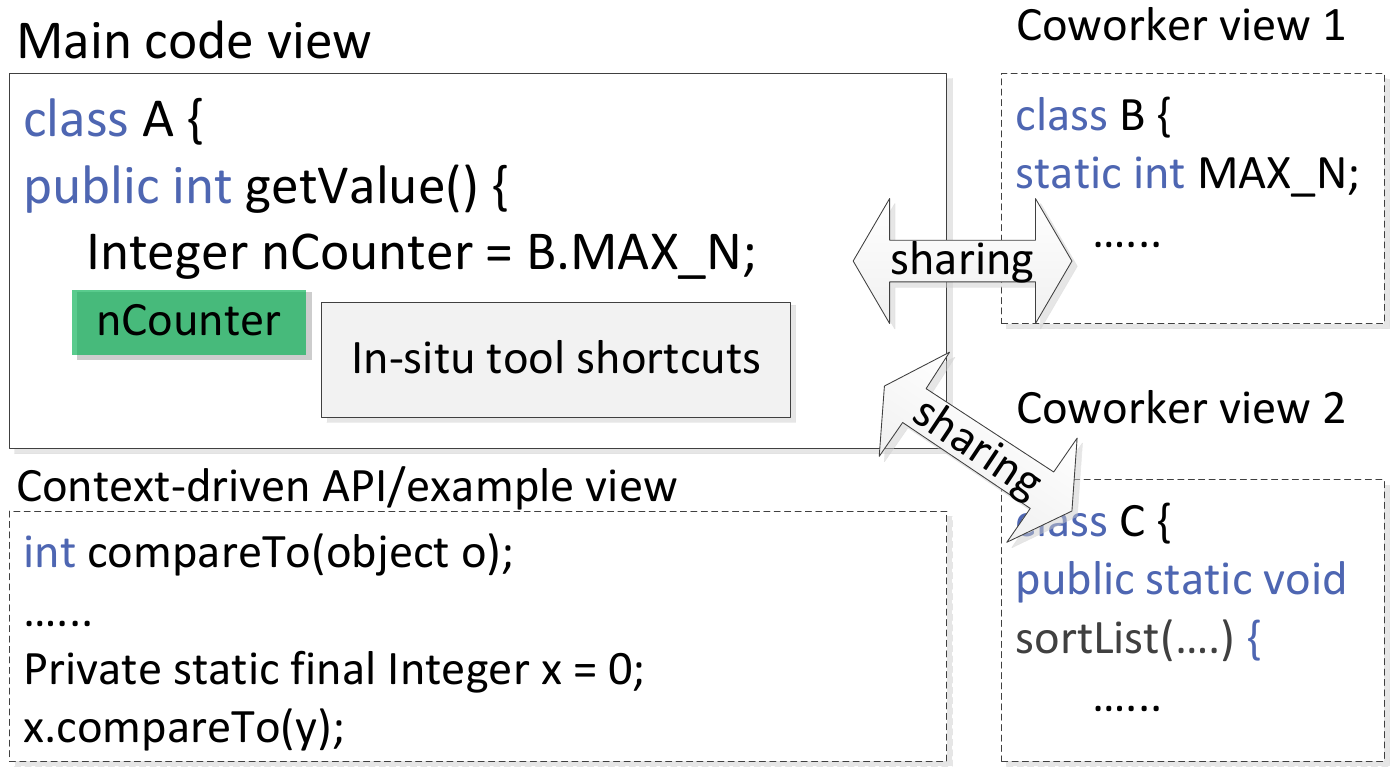}
  %\vspace{-2pt}
  %\caption{Code editing interface that helps reduce context switches of developers between coding and getting aids.}
  \caption{A new interface design that helps reduce context switches of developers between coding and getting aids, and supports close collaborations among coworkers in software development teams.}
  \label{fig:interface}
  %\vspace{-6pt}
\end{figure*}
\vspace{0pt}
Developers spend most of their time on their code for adding new features,
making changes, debugging, and code comprehension~\cite{latoza2006maintaining}.
When doing these tasks, developers often need also external assistances integrated in
their workflow (e.g., automatic code completion~\cite{omar2012active}) which
facilitate their development efficiency. To meet such needs, a tentative framework could incorporate three
interface design features to further reduce workflow transitions of developers when
they are working around their code.

Figure~\ref{fig:interface} gives an overview of these design features.
Aside the traditional coding view, there are a few other co-worker views that assist with 
communication and collaboration tasks typical in team development scenarios; at the bottom, the
context-driven API/example view attempts to provide code examples that 
are recommended based on current coding context to assists programmers with using APIs of which 
usages are not familiar to them; finally, the in-situ interface shown in the main code view 
illustrates the design of porting convenience invocation shortcuts, which are mostly 
spread over varying places in existing IDEs, to the current focus of editing.

\subsection{Context-driven API/example View}
While coding, programmers often have questions
about the usage of some third-party functionalities or features~\cite{latoza2010developers}.
And while implementing a feature, they face hard questions concerning which functions or objects
they should pick~\cite{latoza2010hard}. To some extent, these questions can be reduced to the needs
for getting function usage information and, even further, illustrations of that usage with example
code. Active code completion~\cite{omar2012active} via API menus already helps developers better than using separate
API browsing views, yet it may not be sufficient as developers have to navigate through possibly long API lists (and
hovering on each one to get the function prototype or API documentation on a floating window as seen in
ECLIPSE~\cite{eclipse}), which could potentially break their mental model focusing on the programming logic.

It is plausible to, optionally, put such assistance back into a separate view but closely
connected to the main code editing view (as shown at the bottom of Figure~\ref{fig:interface}), where usage
information of relevant APIs is display on demand based on the current context of object accesses or function calls.
Importantly, all relevant APIs are ranked according to their frequency of being used recently at the default mode.
Similar solutions have actually been explored previously in a more general sense from a perspective of the information
foraging theory, with respect to software engineering tasks such as programming and debugging~\cite{lawrance2013programmers}.

A more important reason for providing the option of moving API usage information to a separate view
is the need of combining code examples with the usage. While showing function prototype and/or API documentation
is helpful to developers to fill in arguments, it is more beneficial to show them usage examples thereof with
the usage synopsis. In practice, programmers search code examples with respect to unfamiliar APIs very often,
by using Internet searches, for instance, even preferably over reading API documents.

In this regard, at least three sources of search for such code examples can be taken into account.
The first one is the examples coming with
APIs in their documentation.
%However, many API documentations do not come with useful examples, when the second source can be checked---
Another option is searching in the current code base for relevant examples using context similarity measurement
(e.g., calling context and/or type of the object from which the API would be invoked). The code shown in the API/example view
of Figure~\ref{fig:interface} illustrates the result obtained from this source: When the cursor lies immediately after
the {\tt Integer} object {\tt nCounter}, the view shows candidate API lists applicable to objects of the {\tt Integer} type,
with ones most frequently used recently listed at the top ({\tt compareTo} here), and followed by the code example found
in the current code base. Such examples give an instant and clear demonstration on how to use the relevant APIs.
Finally, an automatic web search, using open search engine programming interface (e.g., Google API), can be initiated with
queries concerning the function usage (e.g., ``strtoul C++ example"). Then relevant content can be extracted and put back
to the API/example view for programmers' reference.

\subsection{Coworker Views}
Another key interface design feature of our framework concerns about the information needs of
developers collaborating in a development team. Previous studies show that in collaborative development one of the
primary information sources for developers is their co-workers~\cite{ko2007information}. In fact, it is very common that
when developers have questions regarding how a function or feature is implemented, they tend to first resort to their
teammate instead of software documentations~\cite{latoza2006maintaining}. To facilitate developers to take
advantages of having co-workers to consult as their information needs arise, it is potentially rewarding
to incorporate a set of coworker views aside
the main code editing view (shown on the right-hand side of Figure~\ref{fig:interface}).

The rationale of introducing these additional views is two-fold. First, developers working in the
same team can easily share their source code in real-time when
necessary. One example case in which this sharing could be useful is when a senior developer coaches a team member in familiarizing him with the team project. Another example can be seen in agile development, where one developer could quickly
prototype his function according to the ongoing implementation of a function being written or debugged by another developer.
As shown in Figure~\ref{fig:interface}, the current developer is coding the method {\tt getValue()} for class {\tt A}, with a reference to the static variable {\tt MAX\_N} of class {\tt B} that is being coded by a coworker. Having the choice of checking
the implementation of a component, developed concurrently by a teammate, on which current coding task is dependent will save a developer's time seeking for the information about that component in other more expensive ways.

Second, such views can enable close collaboration among physically distributed teammates. For instance, a developer who needs one of his teammates to demonstrate how to write or debug a piece of code would readily get the help from such views without moving to
a different seat or office, or resorting to another instant messaging tools. Screen space allowing, such benefits can be even
augmented with multiple coworker views are opened at the same time---allowing close collaborations among multiple developers.

At the first glance, the above interface designs seemingly conflict our goal of reducing context switches by developers, because
those extra views potentially end up with more context switches. However, the overall cost of context switching will be mostly
reduced indeed as the total time developers would spent on getting the information from these views can be much greater without
these integrated views and information. Consider finding the code example for an API again. Without the automatic code reference
shown within the IDE, a developer would have to search online or consult to other sources that are available usually in different
interfaces from the whole IDE (e.g., a different application like web browser).

\subsection{In-situ Interface Elements}
Almost all IDEs today contain a main menu at the top of the entire interface, followed by one or
several rows of tool shortcuts shown as buttons or icons. Although usually those menus or shortcuts
can be situated differently, few of them is tightly incorporated into the working area of
developers where their functionalities will be applied to. For example, there is always a considerable
``visual distance" from the code being focused on by developers and the shortcuts to functions developers
need to utilize on that code.
While the context switches in such situations are not as large as those seen in cases where developers seek coworker
resources without coworker views, such distance could be much reduced. Accordingly, two possible
interface improvements to reduce the unnecessary distance can be investigated.

First, in-situ tool shortcuts can be added to the main code editing view. The presence of such gadgets is
contingent on user actions of marking focus on (e.g., selecting) code elements to which the shortcuts are applicable;
and the composition of the gadgets is determined by the characteristics of the code elements being focused on by developers.
As an example, Figure~\ref{fig:interface} shows, in the main coding view, an ``in-situ tool shortcut bar'' aside the object
{\tt nCounter} when it is selected through double-click and the mouse cursor hovers nearby---the gadget disappears once
the selection is revoked or the cursor moves away the focused object. This is akin to the in-situ formatting toolbar in
Microsoft Office, triggered by double-clicking on a word.

The more important part of this design is the demand-driven
composition of the gadget. To effectively reduce perceptual transitions within the IDE, the in-situ tool gadget should
contain most, if not all, shortcuts to functionalities that developers would possibly use for the focused object. This
decision can be made in reference to developers' common information needs with respect to that object, based on such
criteria as the object type. For instance, for a function identifier in its invocation statement, example shortcuts
would be ``caller list", ``rename", ``declaration" and so on.

Second, the presence and layout of visual components should be demand-driven. As developers usually work on multiple tasks
during their development workflow~\cite{latoza2006maintaining}, they tend to switch among multiple sources of information.
Yet, they can mostly focus on one task or information at a time only. The IDE thus needs to optimize the size and composition
of the particular visual space where a developer has to concentrate on for a particular task, while diminishing the presence
or even phasing out visual components irrelevant to the current task.

For example, when a developer is right in the process of
typing code, visual components, such as the top menu and main tool bar, side panels, and bottom debugging views, becomes irrelevant and thus should automatically disappear so that the main coding view gets its maximal visual space. Some IDEs,
such as Microsoft Visual Studio, has already incorporated similar features (e.g., dockable gadgets), yet relies on user settings
to apply those features. Also, the overall design there does not support automatic adaptation of interface layout and composition to user action and workflow contexts. The dockable gadgets, for instance, can be set
to hide when mouse cursor moves away from them, but the layout and elements of the gadgets do not automatically accommodate developers' information needs varying in the development workflow.
%The next-generation IDEs can be made more intelligent in this regards by means of
In addition, all visual components that are not applicable to the current developer action can phase out from the interface and
come back when they become applicable again. In contrast, most IDEs choose to disable those components while leaving them in
the visual space.

\section{Visualization Design}\label{sec:vis}
\begin{figure*}[htbp]
  %\vspace{-2pt}
  \centering
  \includegraphics[scale=0.75]{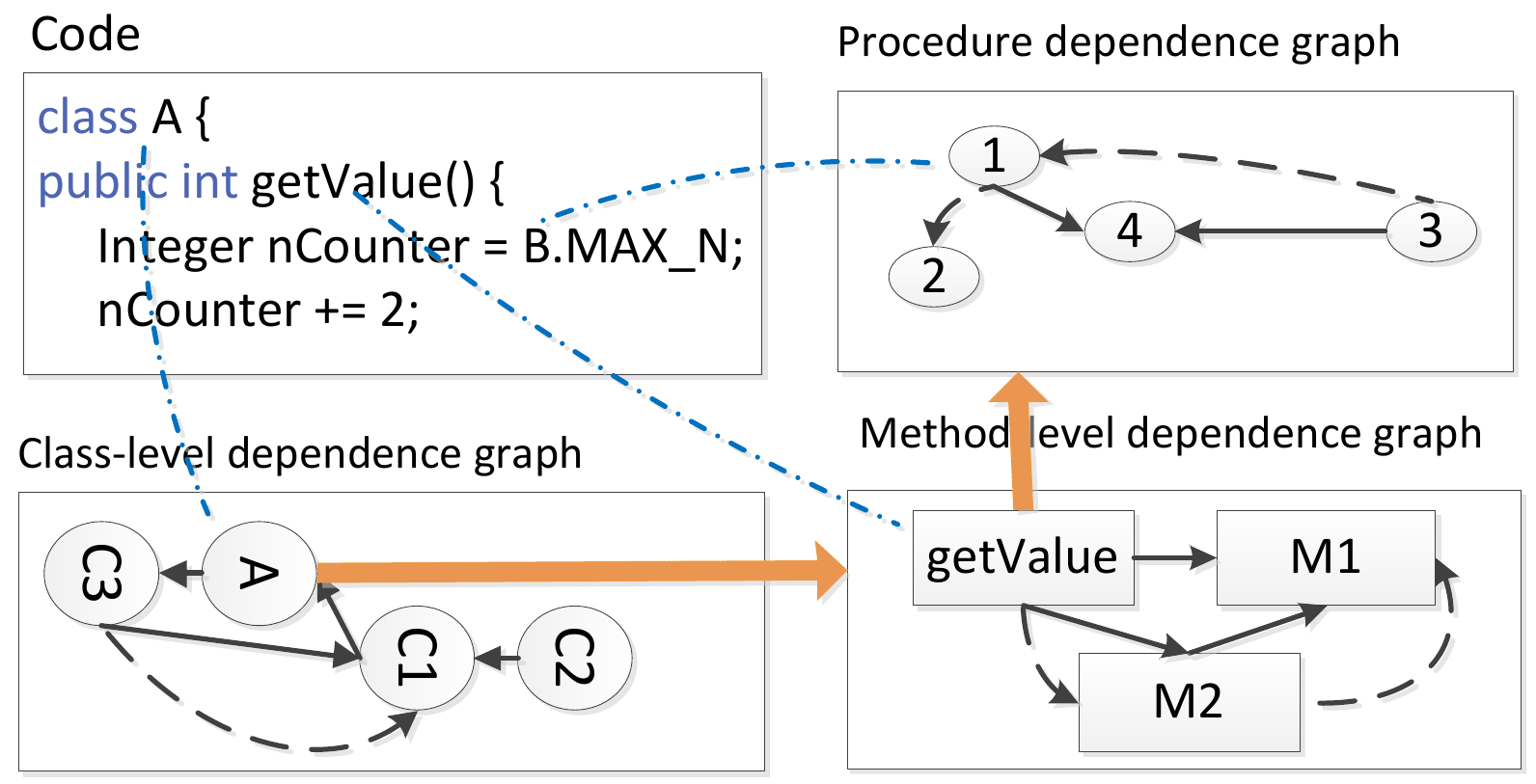}
  \vspace{0pt}
  \caption{Multiple linked visualizations of source code to integrate multiple code information in synergy.}
  \label{fig:vis}
  %\vspace{-2pt}
\end{figure*}
\vspace{0pt}
During software development, programmers gain most of the information they need from the source code they are working on~\cite{latoza2006maintaining,latoza2010hard}. Yet,
they also need information beyond that~\cite{latoza2010developers}, such as those produced by program analysis tools, to obtain better understanding of the software~\cite{zeller2007future}.
Examples of such additional information include call graphs, dependence graphs, and type hierarchies.
While most present IDEs do provide functionalities, via plug-ins for instance, to help developers obtain these
information, they force developers to switch among different visualizations of those information,
potentially leading to expensive workflow interruptions.
An additional issue is that usually these visualizations are separate from each other, without explicit links among them, forcing
developers to maintain an extra mental model linking those information in their mind when manipulating them.
In this context, it is reasonable to leverage multiple linked visualizations of source code,
along with the source code itself, to facilitate the code understanding and navigation for developers.

\subsection{Multiple Visualizations of Source Code}
Not only can information visualization greatly aid data understanding, but also can multiple forms of
information of the same data set be even more helpful (e.g.,~\cite{hanciles1997multiple}). During their
mental workflow for code understanding, which is their primary task~\cite{latoza2006maintaining}, developers
can greatly benefit from multiple visualizations of the source code information besides the textual source code itself.

A relevant proposal would be to utilize multiple visualizations of program source code, which are interconnected underneath the source code,
to enable more effective program understanding. Figure~\ref{fig:vis} illustrates our visualization design feature for the
next-generation IDEs using dependence graphs as the example information representation of source code (there are other
forms of such information of code, such as call graphs and type hierarchies as mentioned above). Surrounding the traditional
code editing view (upper left) that provides the textual representation of source code only, three other satellite views show the dependence graph of the source code at three levels of detail, namely the (statement-level) procedure dependence graph (PDG) (bottom left), method-level dependence graph (MDG) (bottom right), and class-level dependence graph (CDG) (upper right). Different styles of arrows
in the graphs illustrate different types of dependencies. The three dotted lines across the views illustrate the links between the source code and each of these graphs.

Optionally, these visualizations can be selectively or all added to the IDE. A more synergetic design is to
\emph{seamless synthesize} the four views \emph{in one}. In the latter design, instead of showing more than one or
all views at the same time, only one view is visible at a time. The motivation is that, again, while developers can be
greatly benefited from multiple visualizations, they could utilize one of them at one time only. The idea is to switch
among these visualizations by \emph{zooming in/out} operations (shown by two wide arrows in the figure).
As with Google Maps, when developers zoom out from a statement,
they will first switch to the PDG visualization with central points automatically set to that statement; then they can
navigate on the PDG and zoom in from any node thereof back to the source code view. If developers continue to zoom out when
they are in the PDG visualization, they switch to the MDG view where they can also navigate, at the method level, and zoom
in back to the lower levels of details. Switching between MDG and CDG is similar.
Alternatively, zooming in from a method declaration while in the source code view can directly lead the developer to the
MDG visualization, and similarly, zooming in from a class declaration in the source code leads to the CDG directly.

\subsection{Interactions across Linked Visualizations}
The multiple visualizations of source code are also mutually linked, since they all representation the same data (source code).
An additional merit of the multiple linked visualization is more effective interaction. One simple example is that selecting
all lines of code of method can be more easily done by just selecting that method on the MDG visualization view; selecting or
deleting a whole class will see greater benefit in similar ways. Moving code around through interactions on the dependence
graph visualizations would be even more beneficial. For example, a developer can quickly start writing a method by cloning an existing one by copying the corresponding node on the MDG view. When multiple visualizations are shown simultaneously, more interactions can be enabled, such as moving or copying a method from one class to another. Of course, feasible interactions
on the graphical visualizations are subject to feasible automatic source code level operations.

\section{Conclusion and Future Work}
\label{sec:concl}
%\vspace{-2pt}
%We sketched a novel interactive program-analysis pipeline to leverage the merits of information-flow visualization for enhancing the modern IDEs where program analysis tools are integrated. We aim at enabling a seamless transition in such a visual environment
%between traditional coding settings and the visualization interface. Empowered by the three-tier design characteristic of human-centered features, the proposed solution can be realized with the support of an automatic requirement inference at the core of the entire framework.
Today's developers usually deal with multiple tasks simultaneously during their software development process, seeking
variously sources of information for interleaving tasks such as coding, documenting, testing, and debugging. To help
them meet such needs, modern IDEs try to incorporate increasing number of interface elements to provide sources for
meeting those multiple information needs, yet mostly are inclined to actually aggravate the problem of
imposing on the developers the demands of frequently switching among many different contexts.

This paper thus explores in this regard and envisions several interface and interactive visualization design features for enhancing today's IDEs in a way that helps developers meet multiple information needs more efficiently. It illustrated the needs and benefits of incorporating those features in next-generation IDEs with motivating examples.

Beyond what has been tentatively proposed in this paper, there are potentially much more similar features than exemplified to be explored in the future, yet the discussions here could enlighten new lines of research improving programming
interfaces and environments.
% the benefits also illustrate the features with motivating
%examples that support the need of incorporating those features in next-generation IDEs.

An immediate next step would be to develop relevant prototypes of the proposed features on the basis of a popular IDE such as ECLIPSE, and then to evaluate the feasibility and usefulness of such features through user studies with professional developers.

%\section*{Acknowledgment}

\bibliographystyle{splncs03}
\bibliography{paper}

\end{document}